\documentclass[12pt,preprint]{emulateapj}
\usepackage{graphicx}

\shorttitle{Moons as Analogs for Exoplanets}
\shortauthors{Stephen R. Kane et al.}
\slugcomment{Submitted for publication in the Astronomical Journal}

\begin{document}

\title{Solar System Moons as Analogs for Compact Exoplanetary Systems}
\author{
  Stephen R. Kane\altaffilmark{1,2},
  Natalie R. Hinkel\altaffilmark{1,2},
  Sean N. Raymond\altaffilmark{3,4}
}
\email{skane@sfsu.edu}
\altaffiltext{1}{Department of Physics \& Astronomy, San Francisco
  State University, 1600 Holloway Avenue, San Francisco, CA 94132,
  USA}
\altaffiltext{2}{NASA Exoplanet Science Institute, Caltech, MS 100-22,
  770 South Wilson Avenue, Pasadena, CA 91125, USA}
\altaffiltext{3}{CNRS, UMR 5804, Laboratoire d'Astrophysique de
  Bordeaux, 2 rue de l'Observatoire, BP 89, F-33271 Floirac Cedex,
  France}
\altaffiltext{4}{Universit\'e de Bordeaux, Observatoire Aquitain des
  Sciences de l'Univers, 2 rue de l'Observatoire, BP 89, F-33271
  Floirac Cedex, France}


\begin{abstract}

The field of exoplanetary science has experienced a recent surge of
new systems that is largely due to the precision photometry provided
by the Kepler mission. The latest discoveries have included compact
planetary systems in which the orbits of the planets all lie
relatively close to the host star, which presents interesting
challenges in terms of formation and dynamical evolution. The compact
exoplanetary systems are analogous to the moons orbiting the giant
planets in our Solar System, in terms of their relative sizes and
semi-major axes. We present a study that quantifies the scaled sizes
and separations of the Solar System moons with respect to their
hosts. We perform a similar study for a large sample of confirmed
Kepler planets in multi-planet systems. We show that a comparison
between the two samples leads to a similar correlation between their
scaled sizes and separation distributions. The different gradients of
the correlations may be indicative of differences in the formation
and/or long-term dynamics of moon and planetary systems.

\end{abstract}



\section{Introduction}
\label{intro}

The field of explanetary science is constantly expanding and
diversifying. The number of confirmed planets now exceeds 800 but the
number of detected exoplanets may greatly exceed this due to the vast
amount of exoplanet candidates found by the Kepler mission
\citep{bor11a,bor11b,bat13a}. Since most of the multi-planet systems
from the Kepler candidate sample are exoplanets rather than
false-positives \citep{lis12}. Moreover, the number of detected
systems with multiple planets within a relatively small semi-major
axis has increased from both radial velocity (RV) and transit methods.

The compact multi-planet systems are particularly interesting because
their orbital structure and dynamical stability are a challenge to
explain in terms of formation mechanisms. For example, the Kepler-11
system consists of six detected planets, all of which lie within
0.5~AU of the host star \citep{lis11a}. The RV technique was used to
detect the compact system HD~10180, which has seven known exoplanets
(possibly nine according to \citet{tuo12}), four of which have orbital
periods less than that of Mercury \citep{lov11}. There are indications
of trends between the relative radii of planets within a given
multi-planet system \citep{cia13}. A particular trait of compact
multi-planet systems is that they bear a close resemblance to the
system of moons harbored by the giant planets of our own Solar System
when the sizes are placed in units of the host, be that a star or a
planet. This was briefly noted in the case of the Kepler-42 system,
which consists of a low-mass star with three exoplanets \citep{mui12}
and closely resembles the Galilean system of moons.

We present an analysis and comparison of both Solar System moons and
Kepler compact multi-planet systems. We perform this analysis by
scaling the radii and semi-major axis of each body to the radius of
the host: a giant planet in the case of the Solar System and a Kepler
star in the case of the Kepler systems. Section 2 describes the
analysis of the moons in detail and our power-law fit to the data for
the regular (largest) moons. Section 3 repeats this analysis in the
context of the Kepler multi-planet systems, with application to the
systems Kepler-9, 11, 18, 20, 30, 32, 33, and 42. In Section 4 we
quantify the statistical differences between the two
populations. Section 5 presents calculations of the tidal dissipation
timescales for the moons and planets as an additional diagnostic for
comparison. Section 6 describes the comparison of the two samples and
suggests implications for formation mechanisms and subsequent
dynamical evolution in these environments. In particular, we show that
variations in resonance-trapping mechanisms resulting from the
relative disk masses and compositions are possible causes of
differences in correlation gradients between the two samples. In
Section 7, we briefly comment on how these results could be extended
to brown dwarfs.


\section{Moons within our System}
\label{moon}

Here we discuss the radius and semi-major axis distributions for known
moons within our Solar System. Moons are separated into the two broad
categories of regular and irregular satellites. The former are
generally thought to have formed out of the same protoplanetary disk
that formed the planet. The latter are likely captured smaller bodies,
which tend to have inclined and/or retrograde orbits.

\begin{figure*}
  \begin{center}
    \begin{tabular}{cc}
      \includegraphics[angle=270,width=8.2cm]{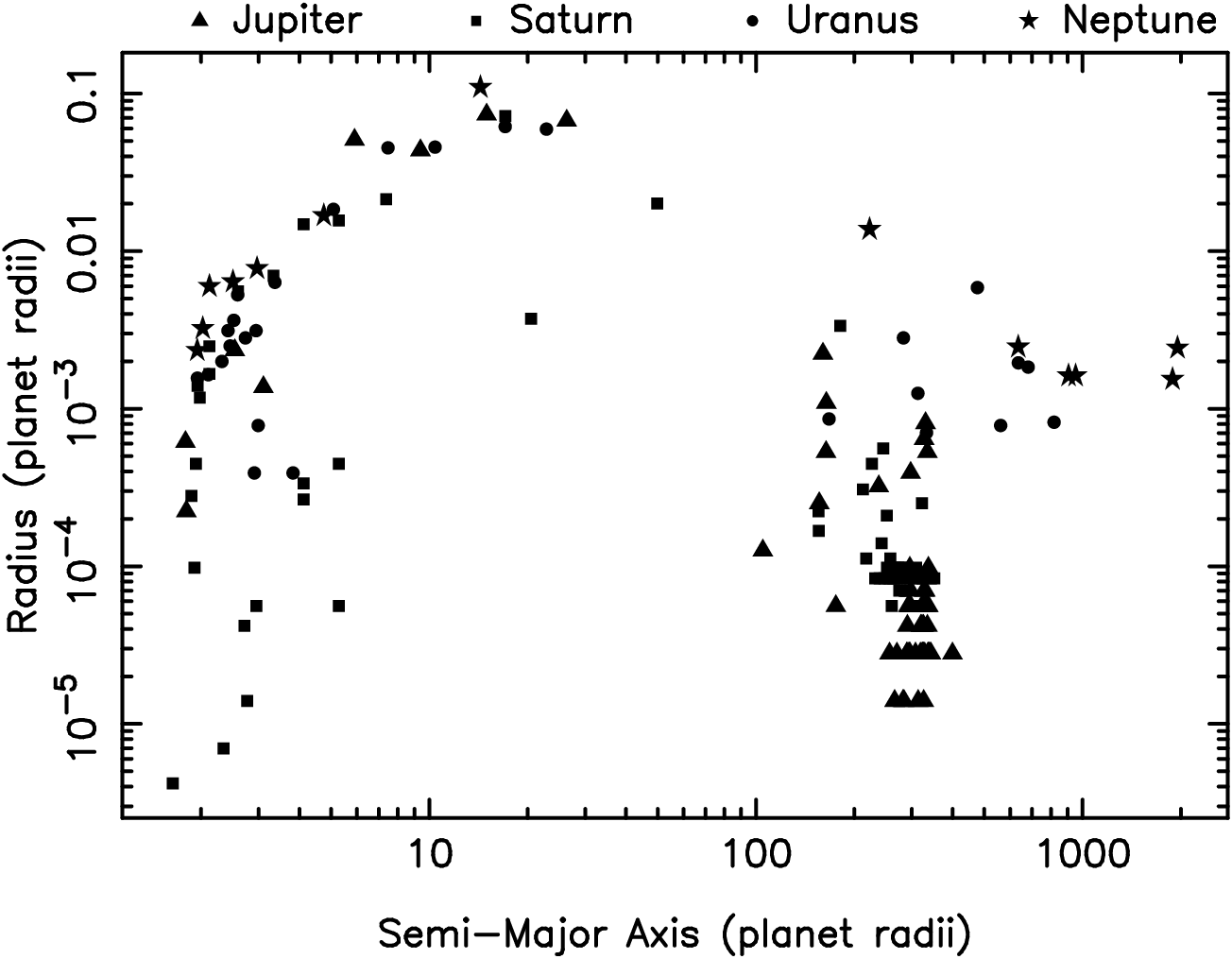} &
      \includegraphics[angle=270,width=8.2cm]{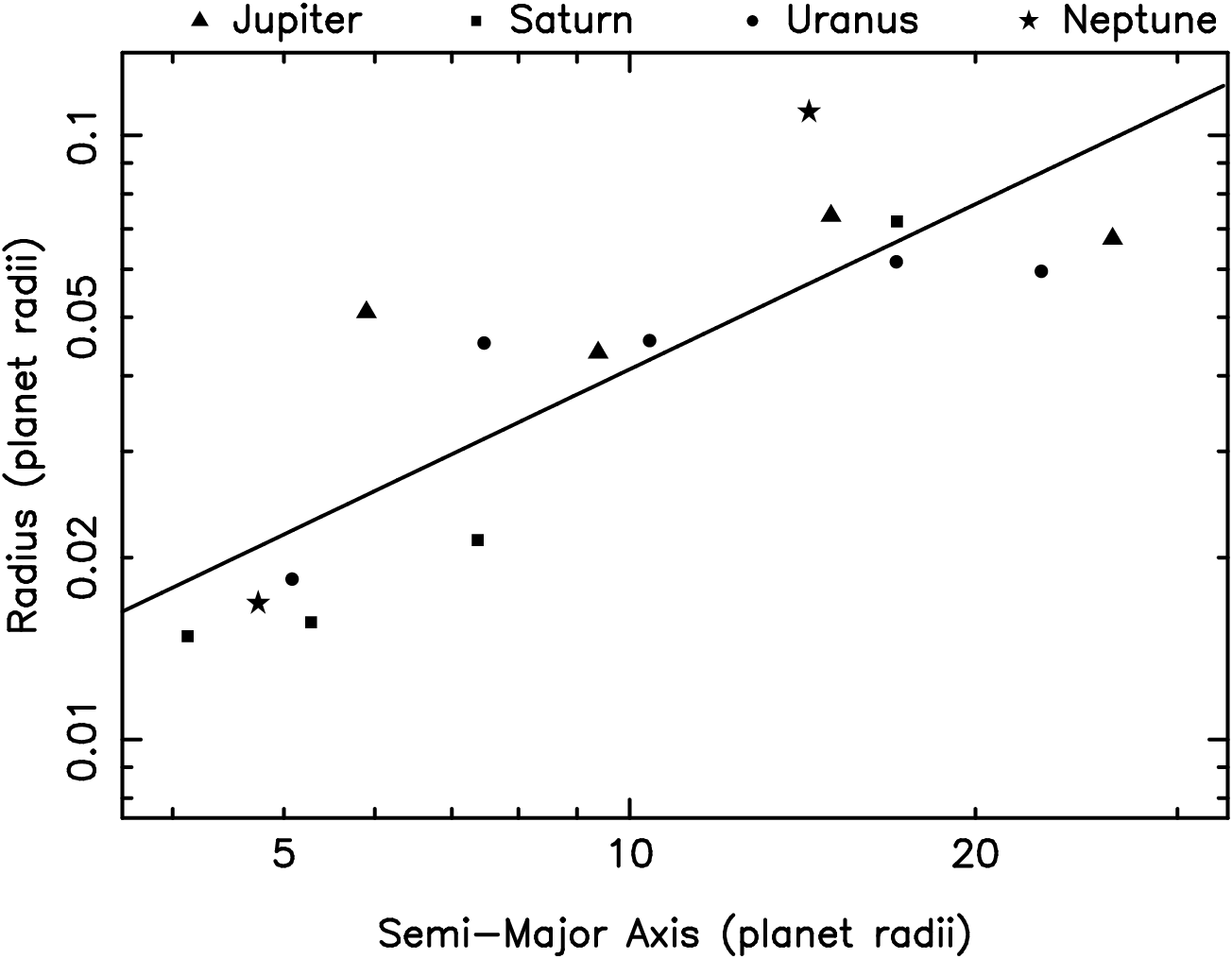}
    \end{tabular}
  \end{center}
  \caption{Left Panel: The dependence of moon radius and semi-major
    axis for all of the giant planet moons of the Solar System. The
    values are scaled in units of the radius of the host planet. Right
    Panel: Zoomed in on the regular moons of the giant planets with a
    power-law fit to the data.}
  \label{moonfig}
\end{figure*}

We include data for 67, 62, 27, and 13 moons for Jupiter, Saturn,
Uranus, and Neptune, respectively. These data were acquired from the
The Giant Planet Satellite and Moon Page\footnote{\tt
  http://www.dtm.ciw.edu/users/sheppard/satellites/} which is curated
by Scott Sheppard \citep{she03,she05,she06}. For each of the 169 Solar
System moons included in this study, we have scaled both the radii and
moon--planet separation to the radius of the host planet. The
resulting scaled radii and semi-major axes for the moons of all four
giant planets are shown in the left panel of Figure \ref{moonfig}.

It is striking in the left panel of Figure \ref{moonfig} how the moons
follow the same distribution for all four of the giant planets. For
radii less than 0.01 of the planet radius, the distribution becomes
dominated by irregular satellites with a range of inclination values
and a mix of prograde and retrograde orbits
\citep{fro11,gas11,jew07}. Those moons beyond 100 planetary radii are
also dominated by irregular satellites whose large separations from
the host planet are artifacts of their capture scenarios. The
Neptunian moon system diverges from the pattern at the outer edge,
likely due to interactions with and captures of Kuiper Belt Objects
\citep{daw12,lev08}. It has been suggested that Triton is a captured
body, possibly from a Kuiper Belt Object binary \citep{agn06}.  The
distribution of irregular satellites close to the planets are
dominated by the shepherd moons of Saturn. The anomalous Saturn moon
near the center of the plot is Hyperion, which is in 4:3 resonance
with Titan \citep{col74,pea78}.

The other major feature of this distribution is that exhibited by the
large moons inside of 30 planet radii. There is a clear correlation of
increasing radius with increasing moon--planet separation. The right
panel of Figure \ref{moonfig} is a zoom of this region which shows the
radius-separation trend more clearly. Table \ref{moontab} contains the
complete list of moons which fall within 30 planet radii, along with
their absolute and scaled semi-major axes and radii. We fit a
power-law to these data of the form $y = c_1 x^{c_2}$ where $c_1 =
0.00507$, $c_2 = 0.90777$, and $c_2$ describes the slope of the
fit. The primary outlier from the fit to the data is the moon Triton
which, as mentioned above, is likely a captured moon. In the next
section we perform a smilar analysis for Kepler multi-planet systems
to investigate the similarity of the correlation.

\begin{deluxetable}{lcccc}
  \tablecolumns{5}
  \tablewidth{0pc}
  \tablecaption{\label{moontab} Solar System Moons Separations and Radii}
  \tablehead{
    \colhead{Moon} &
    \colhead{$a$} &
    \colhead{$R_m$} &
    \colhead{$a$} &
    \colhead{$R_m$} \\
    \colhead{} &
    \colhead{(AU)} &
    \colhead{($R_\oplus$)} &
    \multicolumn{2}{c}{(Host planet radii)}
  }
  \startdata
 \emph{Jupiter} & & & & \\
\ Io       & 0.0028 & 0.572 &   5.900 & 0.051 \\
\ Europa   & 0.0045 & 0.490 &   9.387 & 0.044 \\
\ Ganymede & 0.0072 & 0.826 &  14.972 & 0.074 \\
\ Callisto & 0.0126 & 0.757 &  26.334 & 0.067 \\
 \emph{Saturn} & & & & \\
\ Tethys   & 0.0020 & 0.166 &   4.122 & 0.015 \\
\ Dione    & 0.0025 & 0.175 &   5.279 & 0.016 \\
\ Rhea     & 0.0035 & 0.240 &   7.373 & 0.021 \\
\ Titan    & 0.0082 & 0.808 &  17.091 & 0.072 \\
 \emph{Uranus} & & & & \\
\ Miranda  & 0.0009 & 0.074 &   5.082 & 0.018 \\
\ Ariel    & 0.0013 & 0.182 &   7.469 & 0.045 \\
\ Umbriel  & 0.0018 & 0.183 &  10.407 & 0.046 \\
\ Titania  & 0.0029 & 0.248 &  17.070 & 0.062 \\
\ Oberon   & 0.0039 & 0.239 &  22.830 & 0.060 \\
 \emph{Neptune} & & & & \\
\ Proteus  & 0.0008 & 0.065 &   4.749 & 0.017 \\
\ Triton   & 0.0024 & 0.425 &  14.327 & 0.109
  \enddata
\end{deluxetable}


\section{Compact Multi-Planet Systems}
\label{kepler}

The Kepler mission has numerous multi-planet systems which have been
confirmed using a combination of RV follow-up observations and Transit
Timing Variations (TTV). Here we use the published data for eight
confirmed Kepler multi-planet systems which have three or more planets
in the system: Kepler-9 \citep{hol10,tor11}, Kepler-11
\citep{lis11a,lis13}, Kepler-18 \citep{coc11}, Kepler-20
\citep{fre12,gau12}, Kepler-30 \citep{san12}, Kepler-32
\citep{fab12,swi13}, Kepler-33 \citep{lis12}, and Kepler-42
\citep{mui12}. Using the confirmed multi-planet systems allows for a
more robust analysis since the radii of the host stars are better
characterized than those for the Kepler candidates. We also impose a
scaled semi-major axis cut-off of 120 host radii, beyond which the
systems no longer are compact and comparisons with the Solar System
moons becomes less valid.

\begin{figure}
  \includegraphics[angle=270,width=8.2cm]{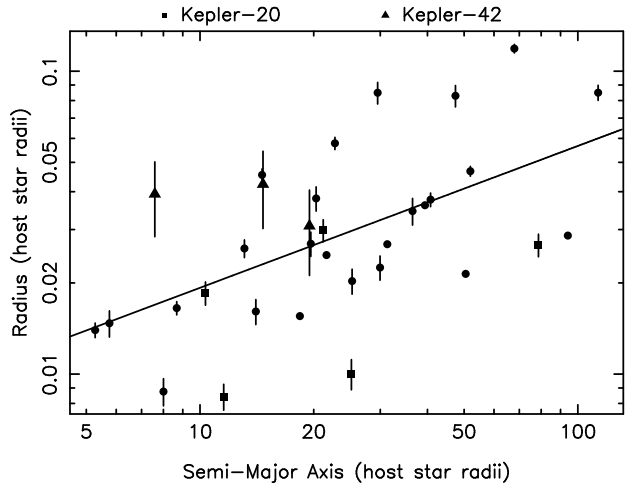}
  \caption{The dependence of planet radius and semi-major axis for the
    planets in eight of the Kepler multi-planet systems. The values
    are scaled in units of the radius of the host star. Also shown is
    a power-law fit to the data.}
  \label{keplerfig}
\end{figure}

As for the Solar System moons described in Section \ref{moon}, we
scale the radii and star--planet separations by the radius of the host
star in each system, shown in Figure \ref{keplerfig} with scaled radii
uncertainties. The Kepler-20 system is specifically marked on the plot
and is unusual because the two smaller planets are staggered between
the three larger planets. Also, the relatively small planets in the
Kepler-42 system have significantly larger radii uncertainties than
those planets in the other systems of this sample. The absolute and
scaled values for all of the Kepler planets included in this study are
tabulated in Table \ref{keplertab}.

\begin{deluxetable}{lcccc}
  \tablecolumns{5}
  \tablewidth{0pc}
  \tablecaption{\label{keplertab} Kepler Planet Separations and Radii}
  \tablehead{
    \colhead{Planet} &
    \colhead{$a$} &
    \colhead{$R_p$} &
    \colhead{$a$} &
    \colhead{$R_p$} \\
    \colhead{} &
    \colhead{(AU)} &
    \colhead{($R_\oplus$)} &
    \multicolumn{2}{c}{(Host star radii)}
  }
  \startdata
 \emph{Kepler-9} & & & & \\
\ b & 0.140 & 9.448 &  29.523 & 0.085 \\
\ c & 0.225 & 9.235 &  47.447 & 0.083 \\
\ d & 0.027 & 1.638 &   5.757 & 0.015 \\
 \emph{Kepler-11} & & & & \\
\ b & 0.091 & 1.807 &  18.379 & 0.016 \\
\ c & 0.107 & 2.873 &  21.610 & 0.025 \\
\ d & 0.155 & 3.120 &  31.305 & 0.027 \\
\ e & 0.195 & 4.197 &  39.383 & 0.036 \\
\ f & 0.250 & 2.491 &  50.492 & 0.021 \\
\ g & 0.466 & 3.333 &  94.116 & 0.029 \\
 \emph{Kepler-18} & & & & \\
\ b & 0.045 & 1.997 &   8.678 & 0.017 \\
\ c & 0.075 & 5.499 &  14.598 & 0.045 \\
\ d & 0.117 & 6.991 &  22.752 & 0.058 \\
 \emph{Kepler-20} & & & & \\
\ b & 0.045 & 1.908 &  10.338 & 0.019 \\
\ c & 0.093 & 3.075 &  21.190 & 0.030 \\
\ d & 0.345 & 2.749 &  78.678 & 0.027 \\
\ e & 0.051 & 0.869 &  11.552 & 0.008 \\
\ f & 0.110 & 1.032 &  25.155 & 0.010 \\
 \emph{Kepler-30} & & & & \\
\ b & 0.180 & 3.905 &  40.755 & 0.038 \\
\ c & 0.300 &12.310 &  67.924 & 0.119 \\
\ d & 0.500 & 8.809 & 113.207 & 0.085 \\
 \emph{Kepler-32} & & & & \\
\ b & 0.050 & 2.199 &  20.292 & 0.038 \\
\ c & 0.090 & 1.997 &  36.525 & 0.035 \\
\ d & 0.128 & 2.704 &  51.947 & 0.047 \\
\ e & 0.032 & 1.504 &  13.109 & 0.026 \\
\ f & 0.013 & 0.808 &   5.276 & 0.014 \\
 \emph{Kepler-33} & & & & \\
\ b & 0.068 & 1.739 &   8.001 & 0.009 \\
\ c & 0.119 & 3.198 &  14.052 & 0.016 \\
\ d & 0.166 & 5.353 &  19.642 & 0.027 \\
\ e & 0.214 & 4.029 &  25.268 & 0.020 \\
\ f & 0.254 & 4.465 &  29.960 & 0.022 \\
 \emph{Kepler-42} & & & & \\
\ b & 0.012 & 0.786 &  14.677 & 0.042 \\
\ c & 0.006 & 0.729 &   7.592 & 0.039 \\
\ d & 0.015 & 0.572 &  19.485 & 0.031
  \enddata
\end{deluxetable}

The range of planet sizes included in our sample varies from
0.572~$R_\oplus$ to 12.31~$R_\oplus$. The radius distribution has a
mean of 3.6~$R_\oplus$ and a standard deviation of
2.9~$R_\oplus$. Part of the reason for the relatively large scatter is
the range of host star properties. Stellar masses are generally not
well determined for the Kepler stars, but the radii range from
0.17~$R_\odot$ to 1.82~$R_\odot$ for Kepler-42 and Kepler-33,
respectively. Despite the dramatic differences in the host stars
however, there is still a visible upward trend to the radius
relationship as a function of semi-major axis. The power-law fit to
the data is once again of the form $y = c_1 x^{c_2}$ where $c_1 =
0.00656$ and $c_2 = 0.46814$. Though still positive, the slope of the
relation is substantially different from that of the Solar System
moons. We discuss this in more detail in Section \ref{formation}.

The lack of data points in the top-left of Figure \ref{keplerfig} is
likely due to a real dearth of planets in that region of
parameter-space, since large planets at short orbital radii are
relatively easy to detect. However, it is worth exploring if the lack
of planets in the bottom-right is due to Kepler incompleteness in that
region since small planets with larger orbital radii are more
challenging to find. The probability of a transit occurring is
dominated by the radius of the star rather than the radius of the
planet \citep{kan08}, so the radius of the planet has a minor
influence on the probability detection bias. The aspect of Kepler
completeness is a complicated issue to quantify since it depends on
the intricacies of the planet detection algorithm with respect to
correlated noise in the data. The Kepler candidates released by
\citet{bat13a} show that transiting planets within the bottom-right
region of Figure \ref{keplerfig} would have been detected if they
existed for the multi-planet systems considered in this paper. In
addition, these multi-planet systems have been studied in far greater
detail than the bulk of the Kepler planet candidates which lends
credence to the superior detection completeness of the multi-planet
systems considered here. Future releases of Kepler multi-planet
systems will provide further understanding of this completeness as the
time-baseline of the Kepler photometry increases.


\section{Statistical Analysis}
\label{statistics}

\begin{figure*}
  \begin{center}
    \begin{tabular}{cc}
      \includegraphics[angle=270,width=8.2cm]{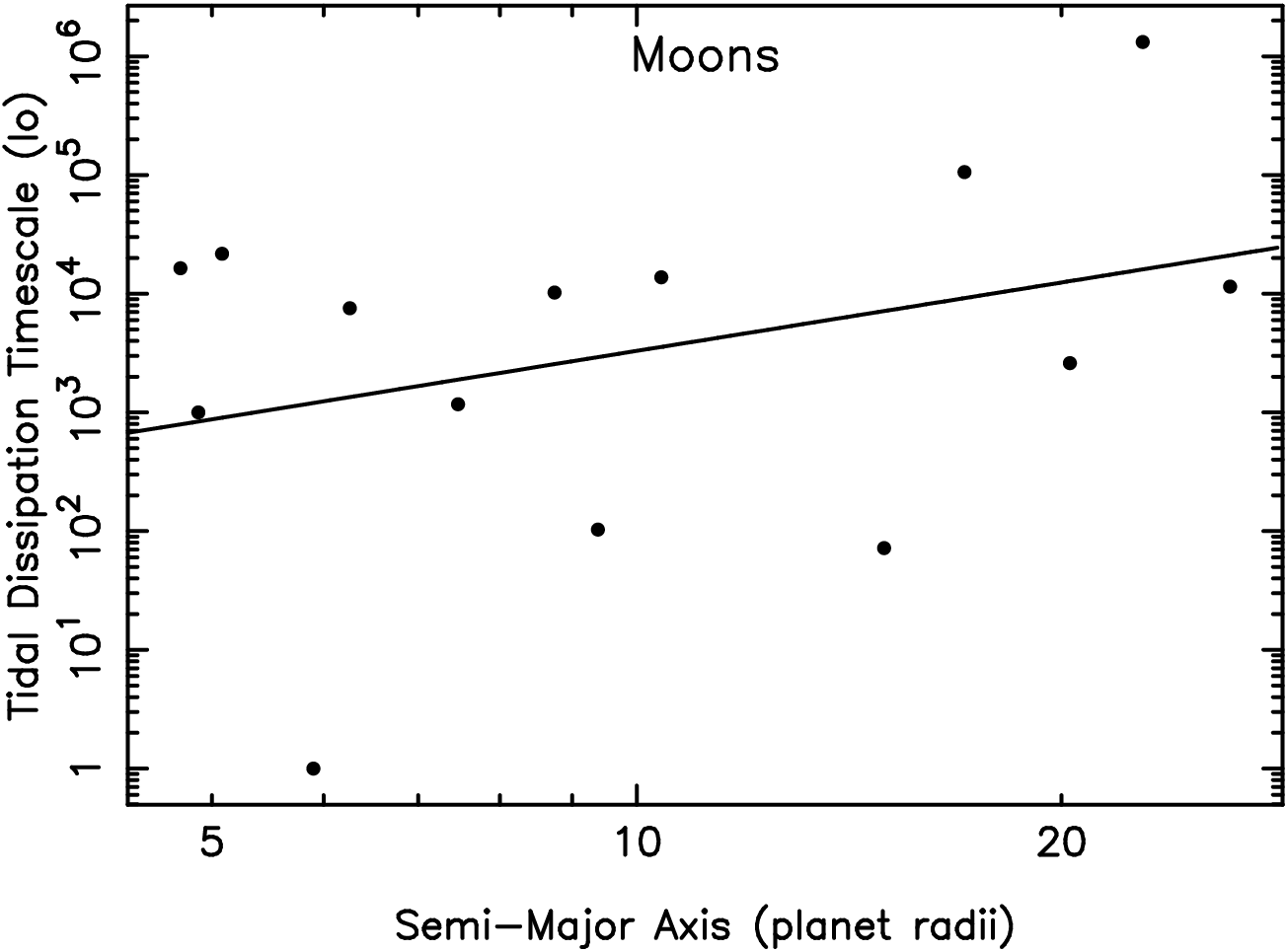} &
      \includegraphics[angle=270,width=8.2cm]{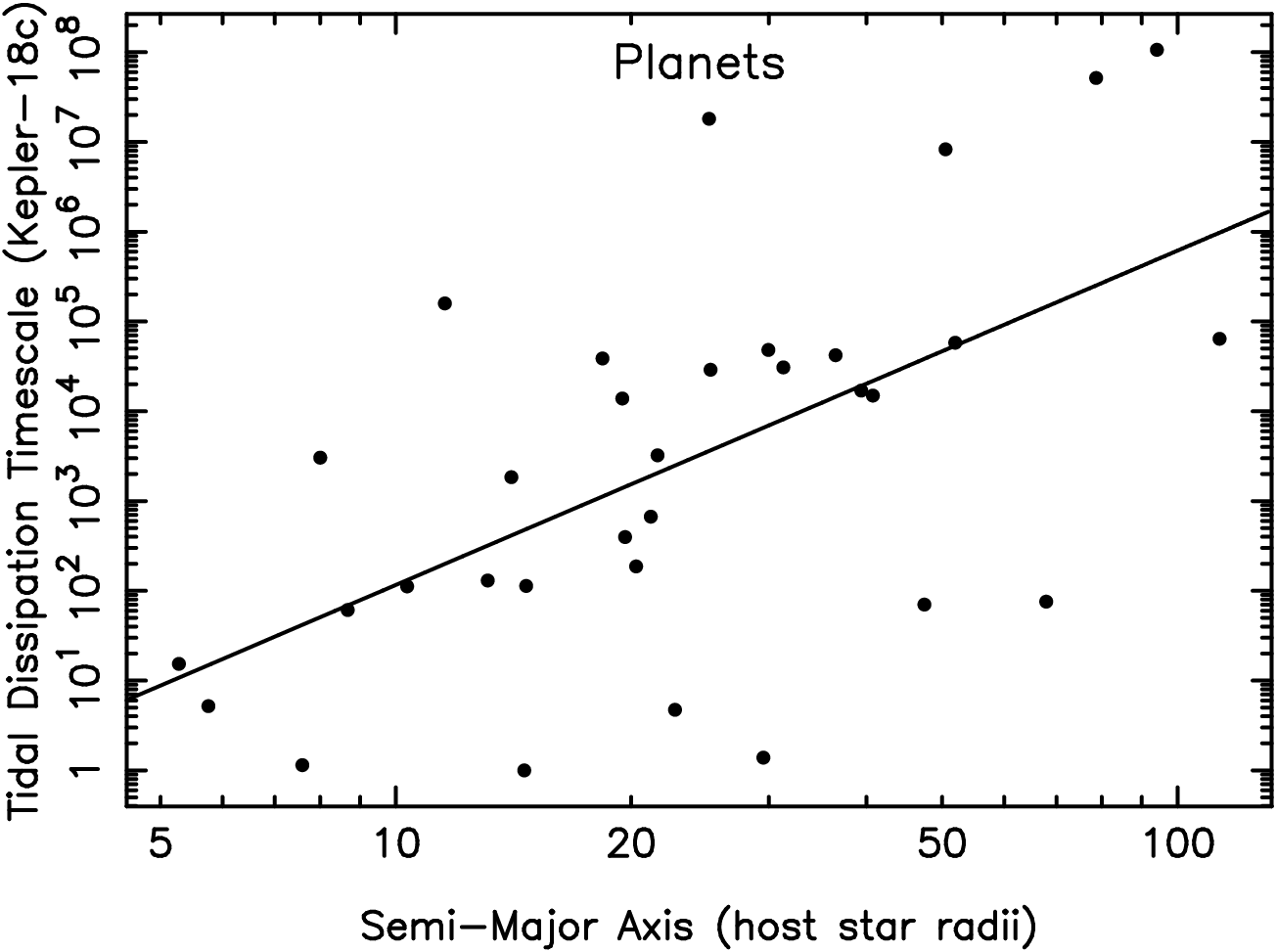}
    \end{tabular}
  \end{center}
  \caption{The dependence of tidal dissipation time-scale on scaled
    semi-major axis for the moon (left panel) and planet (right panel)
    populations. The best-fit power-law is shown in each case.}
  \label{tidalfig}
\end{figure*}

Given the amount of scatter around the power-law fits to the data
shown in Section \ref{moon} (Figure \ref{moonfig}) and in Section
\ref{kepler} (Figure \ref{keplerfig}), we include here statistical
tests to investigate the strength of the apparent correlations. A
robust method to test the correlation lies in the use of the
Spearman's rank correlation coefficient. This delivers a value which
lies in the range $-1 < r_s < 1$, and a corresponding probability of
the null hypothesis (the hypothesis that the two quantities are not
correlated).

Since the data for both the moons and the Kepler planets are best fit
by a power law, we convert the data to a linear scale by using the
logarithm of the values. For the 15 moons which are shown in the right
panel of Figure \ref{moonfig}, the correlation coefficient is $r_s =
0.84$ which means that the data exhibit a strong positive
correlation. The probability that there is no correlation between the
scaled radii and separations is 0.16\%. For the 33 Kepler planets
shown in Figure \ref{keplerfig}, the correlation coefficient is $r_s =
0.53$. This is a positive correlation with a probability of no
correlation of 0.29\%. Even though the scatter in the plot of the
Kepler planets is larger than that for the moons, the probabilities of
the null hypothesis are similar to each other due to the additional
degrees of freedom for the Kepler sample.

To further test the robustness of this result, we constructed a
Monte-Carlo simulation which performs a Fisher-Yates shuffle on the
Kepler data that randomizes the order of the data values. This was
executed 1000 times and in each case the Spearman's rank correlation
coefficient and corresponding probability were recalculated. The
results of this simulation showed that a random ordering of the Kepler
scaled radius values has a mean probability of 50\%, much higher than
that described above. We thus conclude that the correlations shown are
statistically significant.


\section{Tidal Dissipation Timescales}
\label{tidal}

As a further comparison of the moon and planet populations described
in previous sections, we investigate the tidal properties of these
populations. We use the tidal dissipation timescale in the context of
the constant time-lag model \citep{hut81,egg98,lec10}, defined as
\begin{equation}
  T_p = \frac{1}{9} \frac{M_p}{M_\star (M_p + M_\star)}
  \frac{a^8}{R_p^{10}} \frac{1}{\sigma_p}
  \label{tidaleqn}
\end{equation}
where $M_p$ and $R_p$ are the mass and radius of the planet
respectively, $M_\star$ is the mass of the star, $a$ is the semi-major
axis, and $\sigma_p$ is the dissipation factor of the planet
\citep{han10,bol11,bol13}. We estimate the masses of the planets using
the mass-radius relationships of \citet{wei13}. We have calculated
these timescales for each of the objects in the populations shown in
Tables \ref{moontab} and \ref{keplertab}. For the moon population, the
parameters for the star and planet in Equation \ref{tidaleqn} are
replaced with the parameters of the planet and moon respectively. For
each population, we express the dissipation timescales in units of the
object experiencing the highest tidal effects and thus the lowest
tidal dissipation timescale. This object is Io and Kepler-18c for the
moons and Kepler planets respectively.

The results of these calculations are shown in Figure \ref{tidalfig}
for the moon (left panel) and planet (right panel) populations as a
function of their scaled semi-major axes. As was done in Sections
\ref{moon} and \ref{kepler}, we fit a power-law to each distribution
of tidal dissipation timescales of the form $y = c_1 x^{c_2}$. For the
moons, $c_1 = 0.022$ and $c_2 = 3.723$. For the planets, $c_1 =
40.152$ and $c_2 = 1.915$. In this case there is a notable difference
between the two poptulations in that the slope of the relation between
tidal dissipation timescale and scaled semi-major axis for the planets
is approximately twice that for the moons.

Are the differences in tidal timescale distributions of the moons and
planets a result of our assumption of a constant $\sigma_p$? If
$\sigma_p$ is arbitrarily multiplied by 1000 then the two
distributions are more compatible. There is considerable uncertainty
in the dissipation rates of even the best-studied moons, and orders of
magnitude uncertainty for exoplanets. It is reasonable to think that,
given their larger radii, the dissipation rates of planets should be
higher than for moons and their dissipation timescales correspondingly
shorter. However, even if we assume a constant fixed $\sigma_p$ value
for each population, the distributions of the moons and planets are
fit by different slopes. The origin and significance of this
difference is unclear.


\section{Formation Scenarios}
\label{formation}

\begin{figure*}
  \begin{center}
  \includegraphics[angle=270,width=16.0cm]{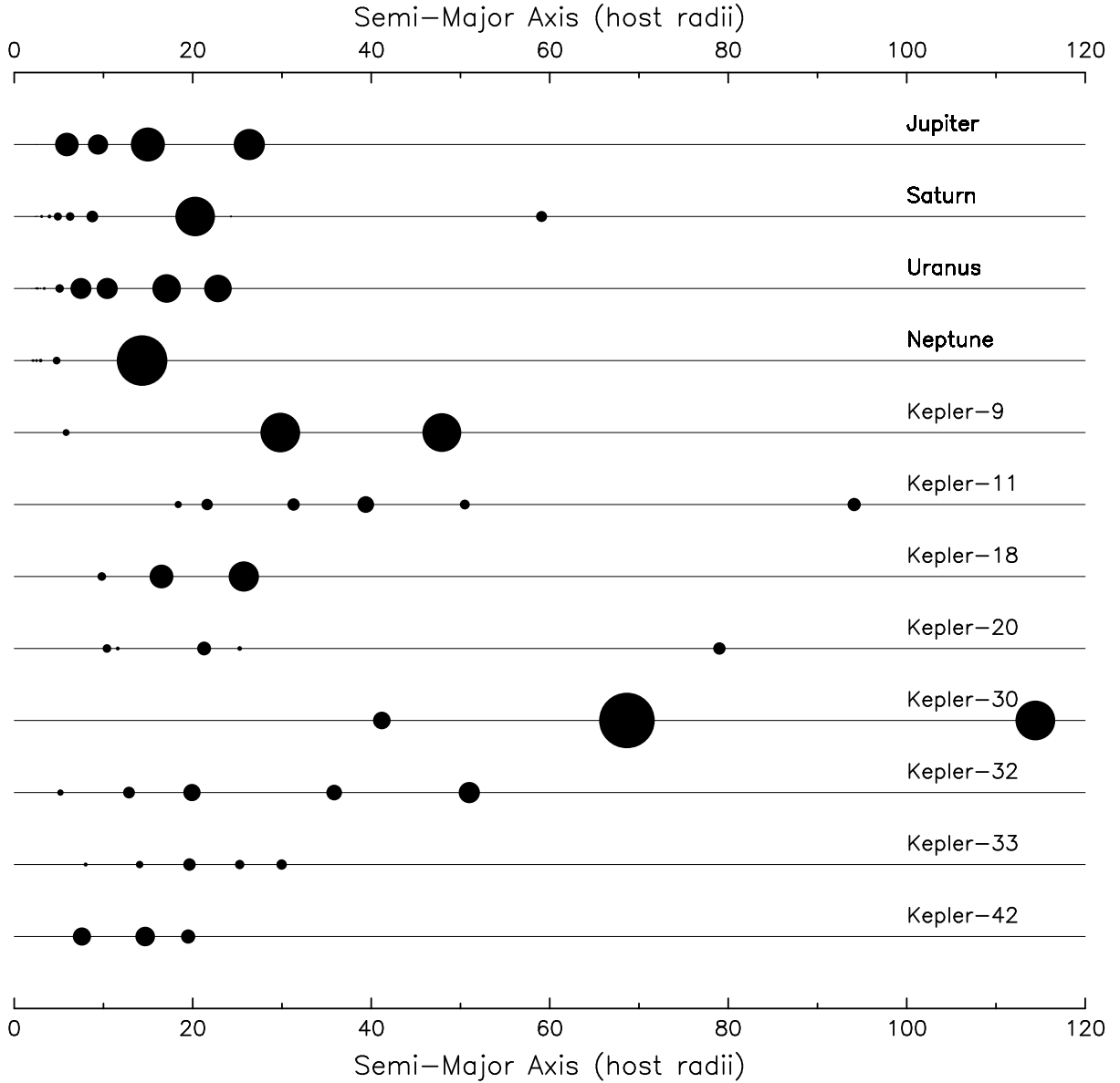}
  \end{center}
  \caption{A visualization of the radii and semi-major axes of the
    Solar System moons and Kepler planets discussed in this paper. The
    radii are all scaled in units of the host, whether the host be a
    planet or a star. Similarly, the separations of the objects from
    their host are scaled in units of the host radii. In
    proportionality to the host, the largest moon is Triton and the
    largest planet is Kepler-30c.}
  \label{radii}
\end{figure*}

A summary of all the Solar System moons and Kepler planets discussed
in this paper are shown in Figure \ref{radii}. The radii are all
scaled in units of the host, which is a giant planet in the case of
the moons and a star in the case of the Kepler planets. Similarly, the
separations of the objects from their host are scaled in units of the
host radii. Two particular bodies stand out in proportionality to the
host: the largest moon (Triton) and the largest planet
(Kepler-30c). It is clear from Figure \ref{radii} that the different
slope of the power-law fits to the data in Sections \ref{moon} and
\ref{kepler} is due to the more diverse range of scaled semi-major
axes of the Kepler planets rather than any significant difference in
scaled radii of the moons/planets. The mean of the scaled semi-major
axes and radii for the moons are 11.5 and 0.048 host radii,
respectively. For the Kepler planets, the mean of the scaled
semi-major axes and radii are 30.9 and 0.035 host radii, respectively.
By comparison, the scaled semi-major axes and radii of the Earth's
moon are 60.3 and 0.273 respectively: significantly removed from the
objects in this study and almost certainly a result of the very
different formation mechanisms involved.

Previous studies have found a host star metallicity correlation with
the presence of giant planets \citep{fis05,ghe10}. Analysis of Kepler
host stars has found that there is no significant correlation of host
star metallicity with the abundance of small exoplanets
\citep{buc12}. A similar analysis on the abundance of chemical
elements in RV host stars also found that the metallicity are not
related to the presence of terrestrial exoplanets
\citep{gon10}. \cite{ram10} further showed the lack of metallicity
correlation in the terrestrial planet hosts by looking specifically at
close solar analogs. Thus the composition of the host is unlikely to
be an indicator of the resulting system properties.

The differences between the Solar System moon and exoplanet
populations are likely due to differences in formation processes. The
compact exoplanetary systems present a particular problem to formation
theories, the various issues of which have been explored by
\citet{ray08a}. In-situ formation requires a very massive close-in
disk \citep{ray08a,chi13}. But in-situ accretion in such a massive
disk would invariably trigger orbital decay due to gas drag and type 1
migration. The inward migration of planetary embryos \citep{ter07}
tends to trap planets in chains of mean motion resonances
\citep{cre08,ogi09}. Resonances can be broken by disk turbulence
\citep{pie11} or on longer timescales by dynamical instabilities
\citep{ter07,mat12} or tidal dissipation \citep{bat13b}. In contrast,
Io, Europa and Ganymede are located in the 4:2:1 Laplace
resonance. This is interpreted as strong evidence that the moons
underwent inward migration either during formation or soon thereafter
\citep{gre87,ogi12,pea02}.

\citet{can06} proposed that the maximum size of moons may be limited
by the migration process. In their scenario, larger moons migrate
faster through the circumplanetary disk and collide with the
planet. An alternative proposal by \citet{cri12} suggests that moons
may be produced by the viscous spreading of massive disks inside the
Roche limit analogous to Saturn's rings. In their model, the
outward-migrating portion of the disk -- pushed out in part by tidal
forcing from the giant planet -- naturally coalesces into one or more
satellites with characteristic properties.

One aspect of the formation differences between the two populations
are the materials available during formation, which can affect the
final composition and density of the bodies. A significant role in the
formation process in the Solar System is played by the ``snow line'',
which is the radial distance from the center of a protostellar disk
beyond which water molecules can efficiently condense to form ice
\citep{ina03}. Beyond the snow line, planetary accretion events have
access to much more material in the form of icey volatiles from which
to form substantial cores for rapid gas accretion
\citep{lis87}. Although this concept has largely been developed in the
context of the Solar System, attempts are being made to apply these
ideas to the diverse range of stellar masses and exoplanetary systems
\citep{ken06,ken08}. The regular moons which form in-situ around giant
planets beyond the snow line may form from similar material to that of
the giant planet, resulting in a lower mean density with respect to
the terrestrial planets, whose formation material consists of mostly
refractory elements \citep{cas01,ray09a,bon10}. If indeed there is a
spatial dependence in the exoplanetary abundances then it will be
reflected in the mean densities of the moons of those planets. For
example, the mean of the mean densities for the Solar System
terrestrial planets is 5.0~g/cm$^3$. The mean of the mean densities
for the Galilean moons is 2.6~g/cm$^3$ and is significantly lower as
one looks to the moons of Saturn, Uranus, and Neptune. Most of the
Kepler comfirmed planets have poorly estimated densities since their
masses often only have upper limits.

An additional comparison that may be made between the two populations
is that of mutual inclinations. The mutual inclinations of the Kepler
multi-planet systems has been an emerging topic as more of those
systems are discovered, and have been studied in some detail by such
authors as \citet{lis11b}, \citet{fan12}, and \citet{tre12}. The
compact Kepler systems have very small mutual inclinations by virtue
of their transit detection, and it was shown by \citet{lis11b} that
the inclination dispersion of these systems generally have a mean $<
10\degr$. Of the Solar System moons shown in Table \ref{moontab}, only
two have inclinations larger than $1.0\degr$ with respect to the local
Laplace plane. These are Miranda ($4.338\degr$) which has strong
evidence of significant geological, tidal, and orbital evolution
\citep{tit90}; and Triton ($156.8\degr$) which, as previously
mentioned, was likely captured into its current retrograde orbit. The
strong mutual inclination present in the remainder of the major moons
is similar to that exhibited by the Kepler compact systems. An
equivalent comparison could be made between the eccentricity
distributions between the two populations as an indicator of analogous
formation mechanisms. However, although the eccentricity distribution
of Kepler planets has been investigated from a statistical perspective
\citep{moo11,kan12}, determination of the eccentricity for individual
planets from photometry alone is a diffcult exercise \citep{kip08}.

A further aspect to the observed orbital configurations of the two
populations is the allowed regions of orbital stability. Stable orbits
for multi-planet systems are a strong function of the mass of those
planets as well as their respective separations, and thus is far more
restrictive for planets than it is for less massive moons
\citep{cha96,ray08b}. Such stability analysis is able to determine
exclusion zones for the presence of planets in the Habitable Zone
\citep{kop10}. Several studies have been undertaken with respect to
the stability of compact planetary systems which have found that these
compact configurations are not only achievable \citep{smi09,fun10},
but may even be the preferred result of planet formation
\citep{ray09b}. Continued discoveries of compact exoplanetary systems
will reveal if these configurations are greatly influenced by these
stability considerations.


\section{Extrapolation to Brown Dwarfs}
\label{browndwarfs}

This work has focussed on the two populations of planets orbiting
stars and moons orbiting planets. In between these two populations
lies a mass regime which includes potential companions to brown
dwarfs. If the same trend could be applied to such systems then one
may expect to find sub-Earth size planets in relatively short-period
orbits around brown dwarfs. Such planets would have exceptionally high
geometric transit probabilities but a successful transit detection
would be hindered by the relative faintness of the brown dwarfs. A
recent study by \citet{bel13} suggests several prospects for
conducting such a survey including space-based or
longitude-distributed network of ground-based observations. A key
advantage in the observational prospects of these systems is that the
orbital period for close-in planets is small enough to allow
monitoring of a complete orbit to occur within a single night of
ground-based observations. However, the expectation that such systems
would follow a similar trend to those described in this work depends
highly upon their formation mechanisms. Thus the detection of planets
around brown dwarfs would play a key role in providing links between
the planet and moon formation scenarios described in Section
\ref{formation}.


\section{Conclusions}

The planetary detections by the Kepler mission have allowed access to
the radii and mean densities for terrestrial planets in multi-planet
systems. One of the big surprises from these multi-planet systems is
the frequency of planets in compact orbital configurations. The high
occurance implies that compact systems are relatively common and bear
the signature of fundamental formation processes that differ
significantly from our own Solar System. However, we have shown here
that there is a correlation between scaled radii and semi-major axes
for these planets (despite the large range of stellar properties) and
that this correlation bears a strong similarity to the same
correlation for the moons of the Solar System giant planets. The main
difference between the two populations is the gradient of the
correlation which is dominated by the Kepler planets extending to
larger scaled orbital separations. The observed correlations may be
explained by inward migration of the moons or perhaps a difference in
materials present during formation. Thus the different correlation
gradients between the two populations is possibly the result of
differences in resonance-trapping mechanisms between proto-moon disks
around planets and proto-planet disks around stars, which in turn is
due to differences in relative disk masses and composition. However,
since migration occurs for both populations, a difference in disk
viscosity and thus turbulent fluctuations may also play a key role in
the resulting compactness of the systems. This may be tested from the
properties of moons which are detected around exoplanets which are
close to their host stars. Searches for exomoons in the Kepler data
are being undertaken by \citet{kip12}. Although there are no current
capabilities to do so, an eventual comparison of the densities between
Solar System moons and exomoons will be useful to study. If the
density of the exomoons differ substantially from the moons of our
Solar System, then it will shed further light on the formation and
migration mechanisms for both the Solar System moons and the Kepler
compact planetary systems.


\section*{Acknowledgements}

The authors would like to thank Jason Eastman, Gregory Laughlin,
Philip Muirhead, and Jonathan Swift for insightful discussions. Thanks
are also due to the anonymous referee, whose comments improved the
quality of the paper. This research has made use of the NASA Exoplanet
Archive, which is operated by the California Institute of Technology,
under contract with the National Aeronautics and Space Administration
under the Exoplanet Exploration Program. The authors acknowledge
financial support from the National Science Foundation through grant
AST-1109662.



\begin{thebibliography}{}

\bibitem[\protect\citeauthoryear{Agnor \& Hamilton}{2006}]{agn06}
  Agnor, Craig B.; Hamilton, Douglas P. 2006, Nature, 441, 192
\bibitem[\protect\citeauthoryear{Batalha et al.}{2013}]{bat13a}
  Batalha, N.M., et al., 2013, ApJS, 204, 24
\bibitem[\protect\citeauthoryear{Batygin \& Morbidelli}{2013}]{bat13b}
  Batygin, K., Morbidelli, A. 2013, AJ, 145, 1
\bibitem[\protect\citeauthoryear{Belu et al.}{2013}]{bel13}
  Belu, A.R., et al. 2013, ApJ, 768, 125
\bibitem[\protect\citeauthoryear{Bolmont et al.}{2011}]{bol11}
  Bolmont, E., Raymond, S.N., Leconte, J. 2011, A\&A, 535, 94
\bibitem[\protect\citeauthoryear{Bolmont et al.}{2013}]{bol13}
  Bolmont, E., Selsis, F., Raymond, S.N., Leconte, J., Hersant, F.,
  Maurin, A.S., Pericaud, J. 2013, A\&A, in press (arXiv:1304.0459)
\bibitem[\protect\citeauthoryear{Bond et al.}{2010}]{bon10} Bond,
  J.C., Lauretta, D.S., O'Brien, D.P. 2010, Icarus, 205, 321
\bibitem[\protect\citeauthoryear{Borucki et al.}{2011a}]{bor11a}
  Borucki, W.J., et al., 2011, ApJ, 728, 117
\bibitem[\protect\citeauthoryear{Borucki et al.}{2011b}]{bor11b}
  Borucki, W.J., et al., 2011, ApJ, 736, 19
\bibitem[\protect\citeauthoryear{Buchhave et al.}{2012}]{buc12}
  Buchhave, L.A., et al. 2012, Nature, 486, 375
\bibitem[\protect\citeauthoryear{Canup \& Ward}{2006}]{can06} Canup,
  R.M., Ward, W.R. 2006, Nature, 441, 834
\bibitem[\protect\citeauthoryear{Cassen}{2001}]{cas01} Cassen,
  P. 2001, M\&PS, 36, 671
\bibitem[\protect\citeauthoryear{Chambers et al.}{1996}]{cha96}
  Chambers, J.E., Wetherill, G.W., Boss, A.P. 1996, Icarus, 119, 261
\bibitem[\protect\citeauthoryear{Chiang \& Laughlin}{2013}]{chi13}
  Chiang, E., Laughlin, G. 2013, MNRAS, in press (arXiv:1211.1673)
\bibitem[\protect\citeauthoryear{Ciardi et al.}{2013}]{cia13} Ciardi,
  D.R., Fabrycky, D.C., Ford, E.B., Gautier, T.N., Howell, S.B.,
  Lissauer, J.J., Ragozzine, D., Rowe, J.F. 2013, ApJ, 763, 41
\bibitem[\protect\citeauthoryear{Cochran et al.}{2011}]{coc11}
  Cochran, W.D., et al. 2011, ApJS, 197, 7
\bibitem[\protect\citeauthoryear{Colombo et al.}{1974}]{col74}
  Colombo, G., Franklin, F.A., Shapiro, I.I. 1974, AJ, 79, 61
\bibitem[\protect\citeauthoryear{Cresswell \& Nelson}{2008}]{cre08}
  Cresswell, P., Nelson, R.P. 2008, A\&A, 482, 677
\bibitem[\protect\citeauthoryear{Crida \& Charnoz}{2012}]{cri12}
  Crida, A., Charnoz, S. 2012, Science, 338, 1196
\bibitem[\protect\citeauthoryear{Dawson \& Murray-Clay}{2012}]{daw12}
  Dawson, R.I., Murray-Clay, R. 2012, ApJ, 750, 43
\bibitem[\protect\citeauthoryear{Eggleton et al.}{1998}]{egg98}
  Eggleton, P.P., Kiseleva, L.G., Hut, P. 1998, ApJ, 499, 853
\bibitem[\protect\citeauthoryear{Fabrycky et al.}{2012}]{fab12}
  Fabrycky, D.C., et al. 2012, ApJ, 750, 114
\bibitem[\protect\citeauthoryear{Fang \& Margot}{2012}]{fan12} Fang,
  J., Margot, J.-L. 2012, ApJ, 761, 92
\bibitem[\protect\citeauthoryear{Fischer \& Valenti}{2005}]{fis05}
  Fischer, D.A., Valenti, J. 2005, ApJ, 622, 1102
\bibitem[\protect\citeauthoryear{Fressin et al.}{2012}]{fre12}
  Fressin, F., et al. 2012, Nature, 482, 195
\bibitem[\protect\citeauthoryear{Frouard et al.}{2011}]{fro11}
  Frouard, J., Vienne, A., Fouchard, M. 2011, A\&A, 532, 44
\bibitem[\protect\citeauthoryear{Funk et al.}{2010}]{fun10} Funk, B.,
  Wuchterl, G., Schwarz, R., Pilat-Lohinger, E., Eggl, S. 2010, A\&A,
  516, 82
\bibitem[\protect\citeauthoryear{Gaspar et al.}{2011}]{gas11} Gaspar,
  H.S., Winter, O.C., Vieira Neto, E. 2011, MNRAS, 415, 1999
\bibitem[\protect\citeauthoryear{Gautier et al.}{2012}]{gau12}
  Gautier, T.N., et al. 2012, ApJ, 749, 15
\bibitem[\protect\citeauthoryear{Ghezzi et al.}{2010}]{ghe10} Ghezzi,
  L., Cunha, K., Smith, V.V., de Ara\'ujo, F.X., Schuler, S.C., de la
  Reza, R. 2010, ApJ, 720, 1290
\bibitem[\protect\citeauthoryear{Gonz\'alez Hern\'andez et
    al.}{2010}]{gon10} Gonz\'alez Hern\'andez, J.I., Israelian, G.,
  Santos, N.C., Sousa, S., Delgado-Mena, E., Neves, V., Udry, S. 2010,
  ApJ, 720, 1592
\bibitem[\protect\citeauthoryear{Greenberg}{1987}]{gre87} Greenberg,
  R. 1987, Icarus, 70, 334
\bibitem[\protect\citeauthoryear{Hansen}{2010}]{han10} Hansen,
  B.M.S. 2010, ApJ, 723, 285
\bibitem[\protect\citeauthoryear{Holman et al.}{2010}]{hol10} Holman,
  M.J., et al. 2010, Science, 330, 51
\bibitem[\protect\citeauthoryear{Hut}{1981}]{hut81} Hut, P. 1981,
  A\&A, 99, 126
\bibitem[\protect\citeauthoryear{Inaba et al.}{2003}]{ina03} Inaba,
  S., Wetherill, G.W., Ikoma, M. 2003, Icarus, 166, 461
\bibitem[\protect\citeauthoryear{Jewitt \& Haghighipour}{2007}]{jew07}
  Jewitt, D., Haghighipour, N. 2007, ARA\&A, 45, 261
\bibitem[\protect\citeauthoryear{Kane \& von Braun}{2008}]{kan08}
  Kane, S.R., von Braun, K., 2008, ApJ, 689, 492
\bibitem[\protect\citeauthoryear{Kane et al.}{2012}]{kan12} Kane,
  S.R., Ciardi, D.R., Gelino, D.M., von Braun, K. 2012, MNRAS, 425,
  757
\bibitem[\protect\citeauthoryear{Kennedy et al.}{2006}]{ken06}
  Kennedy, G.M., Kenyon, S.J. Bromley, B.C. 2006, ApJ, 650, L139
\bibitem[\protect\citeauthoryear{Kennedy \& Kenyon}{2008}]{ken08}
  Kennedy, G.M., Kenyon, S.J. 2008, ApJ, 673, 502
\bibitem[\protect\citeauthoryear{Kipping}{2008}]{kip08} Kipping,
  D.M. 2008, MNRAS, 389, 1383
\bibitem[\protect\citeauthoryear{Kipping et al.}{2012}]{kip12}
  Kipping, D.M., Bakos, G.\'A., Buchhave, L., Nesvorn\'y, D., Schmitt,
  A. 2012, ApJ, 750, 115
\bibitem[\protect\citeauthoryear{Kopparapu \& Barnes}{2010}]{kop10}
  Kopparapu, R.K., Barnes, R. 2010, ApJ, 716, 1336
\bibitem[\protect\citeauthoryear{Leconte et al.}{2010}]{lec10}
  Leconte, J., Chabrier, G., Baraffe, I., Levrard, B. 2010, A\&A, 516,
  64
\bibitem[\protect\citeauthoryear{Levison et al.}{2008}]{lev08}
  Levison, H.F., Morbidelli, A., Van Laerhoven, C., Gomes, R.,
  Tsiganis, K. 2008, Icarus, 196, 258
\bibitem[\protect\citeauthoryear{Lissauer}{1987}]{lis87} Lissauer,
  J.J. 1987, Icarus, 69, 249
\bibitem[\protect\citeauthoryear{Lissauer et al.}{2011a}]{lis11a}
  Lissauer, J.J., et al. 2011, Nature, 470, 53
\bibitem[\protect\citeauthoryear{Lissauer et al.}{2011b}]{lis11b}
  Lissauer, J.J., et al. 2011, ApJS, 197, 8
\bibitem[\protect\citeauthoryear{Lissauer et al.}{2012}]{lis12}
  Lissauer, J.J., et al. 2012, ApJ, 750, 112
\bibitem[\protect\citeauthoryear{Lissauer et al.}{2013}]{lis13}
  Lissauer, J.J., et al. 2013, ApJ, 770, 131
\bibitem[\protect\citeauthoryear{Lovis et al.}{2011}]{lov11} Lovis,
  C., et al. 2011, A\&A, 528, 112
\bibitem[\protect\citeauthoryear{Matsumoto et al.}{2012}]{mat12}
  Matsumoto, Y., Nagasawa, M., Ida, S. 2012, Icarus, 221, 624
\bibitem[\protect\citeauthoryear{Moorhead et al.}{2011}]{moo11}
  Moorhead, A.V., et al. 2011, ApJS, 197, 1
\bibitem[\protect\citeauthoryear{Muirhead et al.}{2012}]{mui12}
  Muirhead, P.S., et al. 2012, ApJ, 747, 144
\bibitem[\protect\citeauthoryear{Ogihara \& Ida}{2009}]{ogi09}
  Ogihara, M., Ida, S. 2009, ApJ, 699, 824
\bibitem[\protect\citeauthoryear{Ogihara \& Ida}{2012}]{ogi12}
  Ogihara, M., Ida, S. 2012, ApJ, 753, 60
\bibitem[\protect\citeauthoryear{Peale}{1978}]{pea78} Peale,
  S.J. 1978, Icarus, 36, 240
\bibitem[\protect\citeauthoryear{Peale \& Lee}{2002}]{pea02} Peale,
  S.J., Lee, M.H. 2002, Science, 298, 593
\bibitem[\protect\citeauthoryear{Pierens et al.}{2011}]{pie11}
  Pierens, A., Baruteau, C., Hersant, F. 2011, A\&A, 531, 5
\bibitem[\protect\citeauthoryear{Ram\'irez et al.}{2010}]{ram10}
  Ram\'irez, I., Asplund, M., Baumann, P., Mel\'endez, J., Bensby, T.
  2010, A\&A, 521, 33
\bibitem[\protect\citeauthoryear{Raymond et al.}{2008a}]{ray08a}
  Raymond, S.N., Barnes, R., Mandell, A.M. 2008, MNRAS, 384, 663
\bibitem[\protect\citeauthoryear{Raymond et al.}{2008b}]{ray08b}
  Raymond, S.N., Barnes, R., Gorelick, N. 2008, ApJ, 689, 478
\bibitem[\protect\citeauthoryear{Raymond et al.}{2009b}]{ray09a}
  Raymond, S.N., O'Brien, D.P. Morbidelli, A., Kaib, N.A. 2009,
  Icarus, 203, 644
\bibitem[\protect\citeauthoryear{Raymond et al.}{2009a}]{ray09b}
  Raymond, S.N., Barnes, R., Veras, D., Armitage, P.J., Gorelick, N.,
  Greenberg, R. 2009, ApJ, 696, L98
\bibitem[\protect\citeauthoryear{Sanchis-Ojeda et al.}{2012}]{san12}
  Sanchis-Ojeda, R., et al. 2012, Nature, 487, 449
\bibitem[\protect\citeauthoryear{Sheppard \& Jewitt}{2003}]{she03}
  Sheppard, S.S., Jewitt, D.C. 2003, Nature, 423, 261
\bibitem[\protect\citeauthoryear{Sheppard et al.}{2005}]{she05}
  Sheppard, S.S., Jewitt, D., Kleyna, J. 2005, AJ, 129, 518
\bibitem[\protect\citeauthoryear{Sheppard et al.}{2006}]{she06}
  Sheppard, S.S., Jewitt, D., Kleyna, J. 2006, AJ, 132, 171
\bibitem[\protect\citeauthoryear{Smith \& Lissauer}{2009}]{smi09}
  Smith, A.W., Lissauer, J.J. 2009, Icarus, 201, 381
\bibitem[\protect\citeauthoryear{Swift et al.}{2013}]{swi13} Swift,
  J.J., Johnson, J.A., Morton, T.D., Crepp, J.R., Montet, B.T.,
  Fabrycky, D.C., Muirhead, P.S. 2013, ApJ, 764, 105
\bibitem[\protect\citeauthoryear{Terquem \& Papaloizou}{2007}]{ter07}
  Terquem, C., Papaloizou, J.C.B. 2007, ApJ, 654, 1110
\bibitem[\protect\citeauthoryear{Tittemore \& Wisdom}{1990}]{tit90}
  Tittemore, W.C., Wisdom, J. 1990, Icarus, 85, 394
\bibitem[\protect\citeauthoryear{Torres et al.}{2011}]{tor11} Torres,
  G., et al. 2011, ApJ, 727, 24
\bibitem[\protect\citeauthoryear{Tremaine \& Dong}{2012}]{tre12}
  Tremaine, S., Dong, S. 2012, AJ, 143, 94
\bibitem[\protect\citeauthoryear{Tuomi}{2012}]{tuo12} Tuomi, M. 2012,
  A\&A, 543, 52
\bibitem[\protect\citeauthoryear{Weiss et al.}{2013}]{wei13} Weiss,
  L.M., et al. 2013, ApJ, 768, 14

\end{thebibliography}
\end{document}